# Current noise enhancement: channel mixing and possible nonequilibrium phonon backaction in atomic-scale Au junctions

Loah A. Stevens[1], Pavlo Zolotavin[1], Ruoyu Chen[1], and Douglas Natelson[1,2,3]

[1]*Department of Physics and Astronomy, Rice University, 6100 Main St., Houston, TX 77005*

[2]*Department of Electrical and Computer Engineering, Rice University, 6100 Main St., Houston, TX 77005*

[3]*Department of Materials Science and NanoEngineering, Rice University, 6100 Main St., Houston, TX 77005*

**Abstract**

We report measurements of the bias dependence of the Fano factor in ensembles of atomic-scale Au junctions at 77 K. Previous measurements of shot noise at room temperature and low biases have found good agreement of the Fano factor with the expectations of the Landauer- Büttiker formalism, while enhanced Fano factors have been observed at biases of hundreds of mV [R. Chen *et al.*, Sci. Rep. **4**, 4221 (2014)]. We find even stronger enhancement of shot noise at 77 K with an "excess" Fano factor up to ten times the low bias value. We discuss the observed ensemble Fano factor bias dependence in terms of candidate models. The results are most consistent with either a bias-dependent channel mixing picture or a model incorporating noise enhancement due to current-driven, nonequilibrium phonon populations, though a complete theoretical treatment of the latter in the ensemble average limit is needed.

## I. Introduction

Fluctuations about the average current, or current noise, provide additional information about an electronic system beyond that available from the linear response conductance near zero bias, or from *I-V* characteristics. Evaluation of electronic noise can provide information about the dynamical properties of electron transport such as the effective charge of the quasiparticles [1,2] or electronic temperature [3].

In the absence of an applied bias, the equilibrium current noise of a device is given by the Johnson-Nyquist thermal noise [4,5]. Upon the application of a bias, the device is driven into a nonequilibrium steady state of current flow. In this case, additional or "excess" noise is expected, including shot noise, which is attributed to the discrete nature of the electronic charge. Shot noise has constant spectral power density over a broad range of frequencies and is generally proportional to the average current, $I$, across the device [6]. The amplitude of the mean square current fluctuations per unit bandwidth, $S_I$, is commonly expressed for shot noise as $F \times 2eI$ where $F$ is the Fano factor and $e$ is the electronic charge [7].

In mesoscopic systems, transport in the linear regime is often well described by the Landauer-Büttiker model [8]. In this noninteracting approach, conductance, $G$, is defined in terms of the summed transmittances, $\tau_i$, of discrete quantum channels as $G = G_0 \sum_{i=1}^{N} \tau_i$, where $G_0 \equiv 2e^2/h$ is the conductance quantum, and the Fano factor is $F = \frac{\sum_{i=1}^{N} \tau_i (1-\tau_i)}{\sum_{i=1}^{N} \tau_i}$. Suppression of shot noise in fully transmitting channels ($\tau_i = 1, F = 0$) has been successfully verified in atomic-scale junctions at low temperatures [9] and at room temperature [10], as well as in junctions containing molecules [11]. In ensembles of metal junctions, as conductance exceeds 2-3 $G_0$, it is common to see only a partial suppression of shot noise at values close to multiples of $G_0$. This behavior is attributed to channel mixing, and by appropriate analysis, the $\tau_i$, may be estimated from the noise and conductance data [12-14].

In previous experiments on ensembles of atomic-scale gold junctions at room temperature [15], low bias measurements were consistent with the finite-temperature Landauer-Büttiker picture. At source-drain biases above ~150 mV, however, the ensemble-averaged excess noise was enhanced relative to the expected bias dependence of $S_I$ extrapolated from low biases. Quantitative analysis ruled out flicker noise [16] and bulk heating of the electrodes [17] as the likely sources of this increase. Remaining candidate contributions to the noise consistent with the observed bias dependence included modification of $F$ due to electron-phonon interactions (though theoretical treatments generally focus on noise in individual junction configurations, not the ensemble average we report), most consistent with quasi-equilibrated phonon populations [18,19], and local electronic heating from Fermi liquid shear viscosity effects [20,21].

To investigate these alternatives, the present work examines the bias dependence of shot noise in STM-style Au break junctions at 77 K. A large number of successive formation and breaking cycles of an atomic scale contact in an STM-style experimental setup provides information about an average of the system behavior. Each individual junction geometry lasts for only a short period of time. For example, a typical conductance plateau during a breaking cycle at low bias and 77 K lasted on the order 100 ms. By collecting data rapidly during this short duration and over many repetitions, one is less limited by junction instabilities at high biases seen in the mechanical break junction setup [22].

The temperature 77 K was chosen to be low enough to reduce the equilibrium population of the Au optical phonon (~17 meV or 200 K), yet high enough to avoid large contributions from bulk heating of the electrodes. The enhanced temporal stability of these junctions, compared to similar structures at room temperature, drastically reduces the contribution of transient states during atomic rearrangements and enables more detailed quantitative analysis. Again, we observe good agreement with the Landauer-Büttiker model at applied biases below ~180 mV. In this low bias regime, inferred transmittance distributions [14] are consistent with previous studies of channel mixing in Au junctions. As bias is increased, we find a global, weakly conductance-dependent increase in the effective Fano factor. The simplified model previously proposed by Chen *et al*. [15] involving electronic heating strictly local to the junction is not consistent with the observed conductance and bias dependence of the Fano factor. Qualitative similarities between the present results and the room temperature data, despite a nearly four-fold decrease in equilibrium temperature, strongly suggest that interactions with equilibrated phonon populations cannot be responsible for the noise enhancement. Based on the bias dependence of the observed noise enhancement, we conclude that the present results could be explained by a specific scheme of bias-dependent channel mixing, or potentially or by nonequilibrium phonon backaction effects [18,19] if individual junction/single-channel theory predictions of bias dependence are robust when extended to multiple channels and the ensemble average situation.

## II. Methods

A schematic of the experimental setup for measuring high frequency current noise is outlined in Figure 1, which is the same as those described in previous reports by our group, namely Refs. 10, 15, and 23. Our approach is broadly similar to that of Refs. 24 and 25. A square wave is sourced from a function generator to the junction, toggling between 0V and $V_{max}$, at 40 kHz, to measure conductance as junctions are formed and broken (low-frequency part of setup is highlighted in blue in Figure 1). Following the sample is a current-limiting resistance standard, put in place to avoid overloading the current amplifier when the junction is in a shorted, high-conductance state. From the current pre-amplifier, the signal is measured via lock-in amplifier, which is synchronized to the square wave. Finally, the lock-in output is digitized by a data acquisition module (DAQ).

The high-frequency side (red in Figure 1) is used to measure the noise characteristics of the junction. When the bias of the square wave is applied across the junction, both shot noise and Johnson-Nyquist noise are present. The noise signal is transmitted via coaxial cable through a chain of amplifiers and then bandwidth limited to roughly 250-600 MHz. This high-frequency bandwidth is chosen to limit contributions of *1/f* noise. The signal is then fed to a logarithmic power detector whose output is transmitted to both a second synchronized lock-in as well as directly to the DAQ.  The lock-in noise signal is proportional to $S_I(V_{max})$-$S_I(0)$ and is converted to units of $A^2/Hz$ by using the direct measurement from the power detector to the DAQ as a basis for determining absolute noise power.

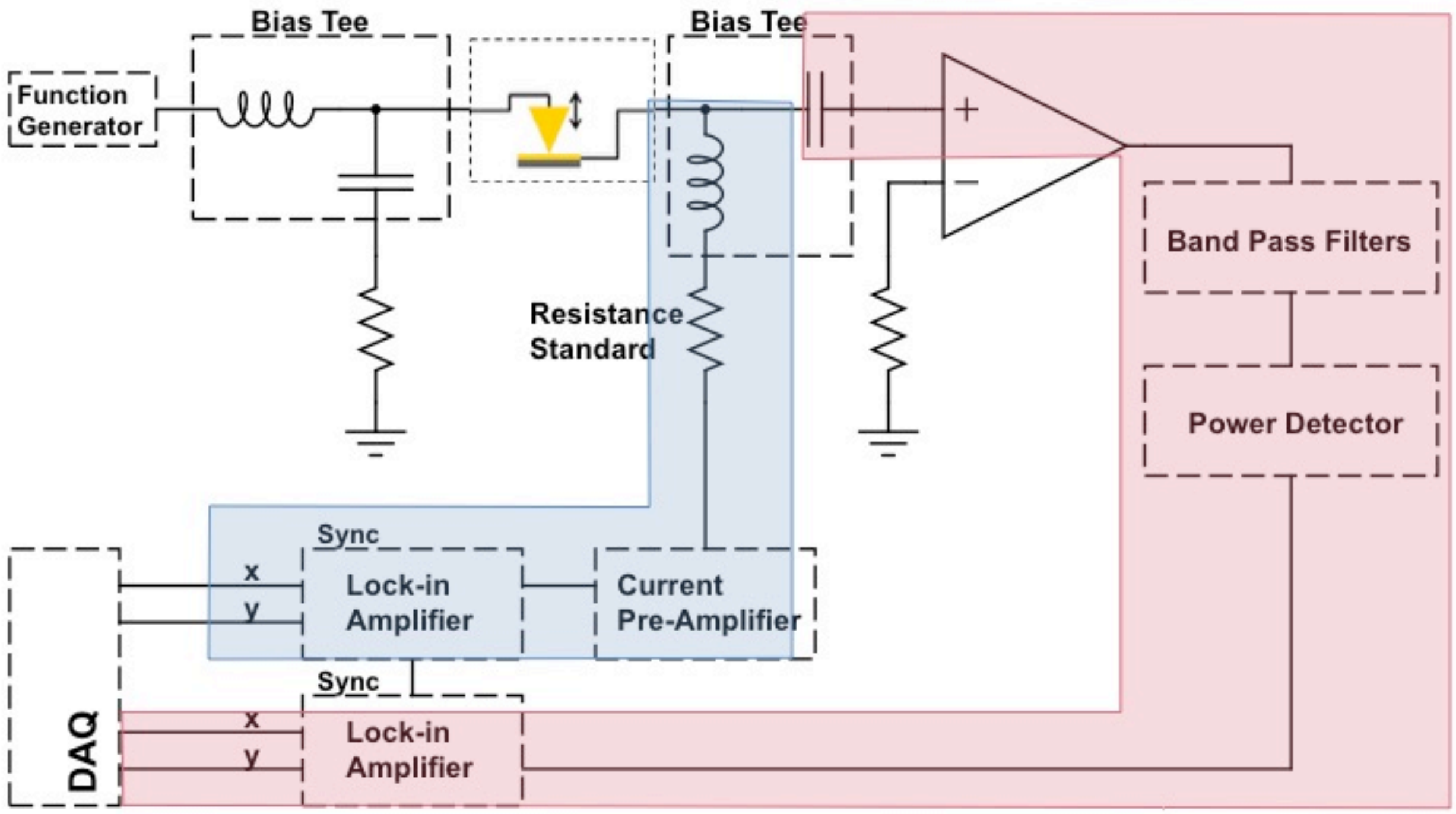

**Figure 1.** Schematic of experimental setup for simultaneous measurement of conductance and RF shot noise.  Device is connected to circuit via the STM-style tip and gold film.  Arrows indicate motion of gold tip.  Low-frequency portion is outlined in blue, and RF subsection is in red (right).

Measurements were conducted in a cryostat-based probe station with attocube piezo positioners and minicoaxial wiring for radio frequency (RF) signals, with substrate and tip temperature set at 77 K via helium exchange gas.  Special care was taken to isolate the system from both external RF signals and mechanical vibrations.  A thick gold wire carved at the end to a sharp tip is continuously moved into and out of contact with a 300 nm thick gold film evaporated onto a silicon substrate.  Over a period of 4.2 seconds, atomic-scale junctions are repeatedly formed and broken between the two gold contacts. The data was sampled at a rate of $2x10^5$ samples/second.

To obtain ensemble averages of noise at a given conductance for a range of biases, around 1000-1500 breaking traces were acquired at 20 different applied biases. Conductance histograms were collected, and the presence of the expected peaks near 1, 2, and 3 $G_0$ with low background was used as figure of merit for each data set.

The advantage of this measurement approach is the ability to acquire such ensemble-averaged noise and conductance data relatively efficiently. We note that a disadvantage is that this method does not permit the simultaneous acquisition of *IV* or differential conductance traces for individual junction configurations.

After collection, the raw data was processed with an automated, multistep procedure to remove transitional points between conductance plateaus, leaving only the noise data collected during stable junction configurations. The first step in the procedure removes data points that are adjacent to the transients in the logarithmic power detector output that occur when the junction resistance changes abruptly during breaking. The readings following a resistance change for which the background noise level exceeds two standard deviations of the open circuit background are removed, as well as the following sixty points, thus ensuring all potentially spurious data surrounding a transient are caught. In the next step, points with conductances that are not within 1% of the 20 preceding readings are removed. The final check ensures all remaining points are within 3% of a neighboring point, leaving only data within stable conductance plateaus. This procedure minimizes any artifacts of stepwise changes in conductance, which can cause speciously high standard deviations of the noise when the junction is undergoing a change in configuration, despite relatively little data in these regions [26].

The noise is averaged over conductance bins of 0.01 $G_0$, chosen to be small to avoid excess averaging. The subtracted background for each applied bias was calculated as the mean of the measured noise at the 10,000 points with the lowest conductance values at that bias (closest approximation to the broken state $G = 0$). This background, when converted via amplifier gains into units of current noise was found to be approximately $1 \times 10^{-24}$ $A^2$/Hz at every applied bias. At finite temperature in the Landauer-Büttiker regime, we expect the shot noise contribution to excess noise to be given by $S_I = 4k_BTG_0\sum_{i=1}^{N}\tau_i^2 + 2eVG_0\coth\left(\frac{eV}{2k_BT}\right)\sum_{i=1}^{N}\tau_i(1-\tau_i)$, which should be linearly dependent on a scaled bias $X = 4k_BTG\left[\frac{eV}{2k_BT}\coth\left(\frac{eV}{2k_BT}\right) - 1\right]$. The ensemble-averaged Fano factor at each conductance bin was then calculated as $F = \frac{S_I(V)-S_I(0)}{X}$, where $S_I(V)$-$S_I(0)$ is the change in noise measured by the lock-in amplifier.

The calculated Fano factor was then plotted as a function of conductance and compared against the minimum Fano factor line, modeled for only one partially open channel at any given conductance. This provides a means of establishing the overall normalization of the $S_I(V)$-$S_I(0)$ signal by using the lowest bias dataset as a reference point.

## III. Results and Analysis

### *A. Low Bias Data*

At low biases, the measured current noise was found to agree well with the expectations of the Landauer-Büttiker picture. Figure 2 shows the data and analysis for a bias voltage of 160 mV applied across the junction and the current-limiting series resistor. The conductance histogram (Figure 2(a)) displays the characteristic peaks near integer values of $G_0$, as expected from past work on Au junctions [10,26,27]. Figure 2(b) is the calculated ensemble-averaged Fano factor as a function of conductance, including the standard error defined as $\frac{\sigma}{\sqrt{N}}$, where σ is the standard deviation of $(S_I(V)$-$S_I(0))/X$ in a given conductance bin and $N$ is the number of data points in the bin. In the range of roughly 0.8-3 $G_0$, the data closely resembles the minimum Fano factor curve (solid black line), the theoretical expectation for the Landauer-Büttiker model when individual transmission channels open sequentially. This relatively clean successive channel opening has been observed previously by others [14,22,28]. Note that some channel mixing is naturally expected at higher conductances, as the number of possible atomic configurations compatible with a given conductance grows rapidly with $G$.

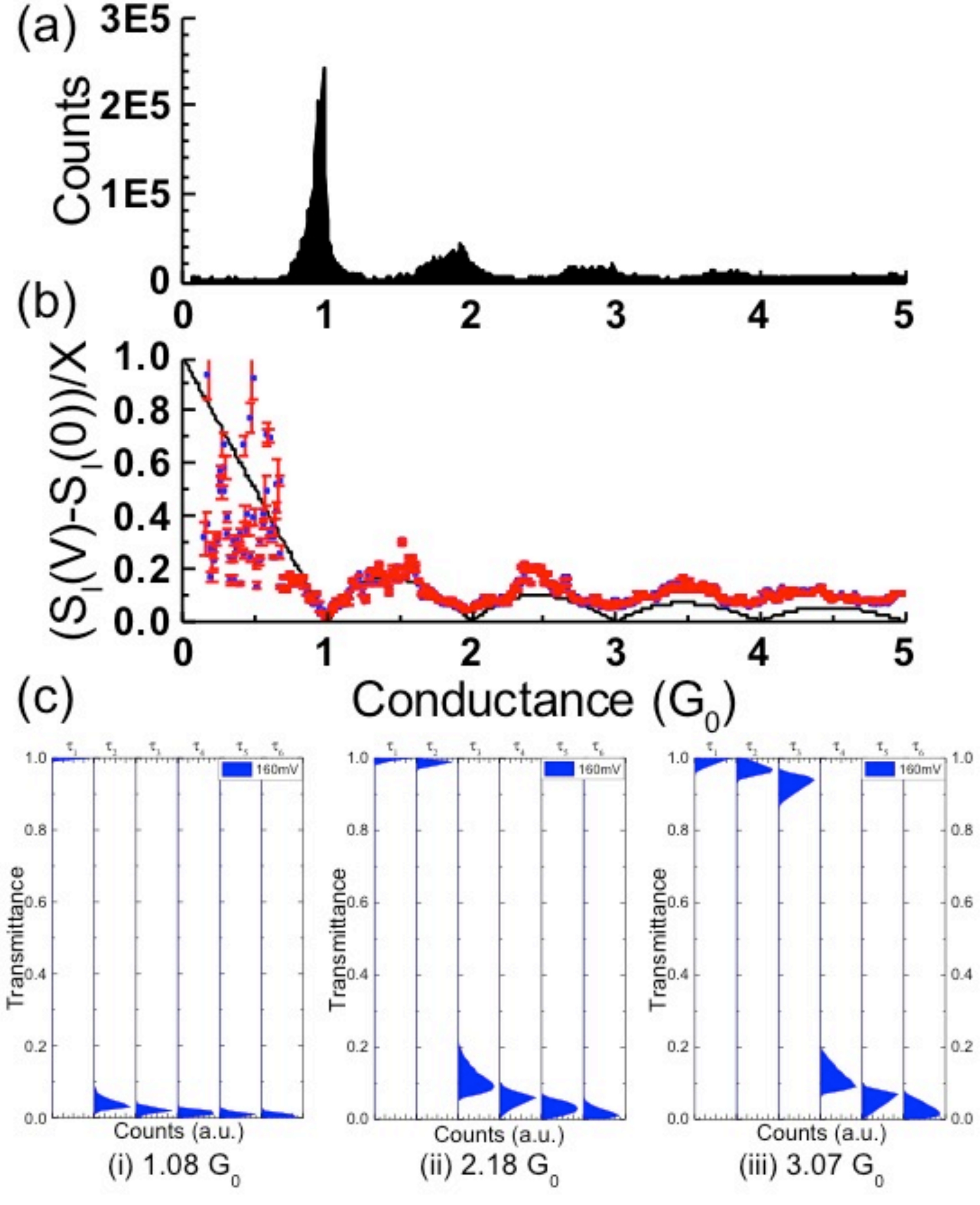

**Figure 2.** **(a)** Conductance histogram and **(b)** ensemble averaged Fano factor as a function of conductance for an applied bias of 160mV. Theoretical minimum Fano factor line is shown in black. **(c)**

Transmittance histograms for $G$ =1.08, 2.18, and 3.07 $G_0$ and Fano factors of 0.07, 0.09, and 0.09, respectively.

Some points are observed in the "forbidden" regime below the theoretical $F(G)$ line at $G < \sim 0.75\ G_0$. Over the course of a breaking cycle, comparatively little data is collected in this conductance range because junction configurations are considerably less stable for $G < 1\ G_0$ and most data points are culled by the automated transient-removal analysis procedure. This small number of counts is apparent in the accompanying histogram for $G < 0.75\ G_0$. Consequently, noise data for conductances below ~ 0.75 $G_0$ were much sparser and more broadly distributed than at higher conductances. The efficiency of RF detection also appears to decrease at low conductances. We attribute the data points in the "forbidden" regime (depicted in Figure 2(b)) to this sparseness and consequent lack of adequate averaging in this conductance region, and to degraded RF detection in this range, and *not* to spin-polarized conduction seen in other recent experiments by Vardimon *et al.* in molecular junctions between ferromagnetic electrodes [29]. Because of these systematic issues in this low conductance limit, we do not use this low conductance data in the following analysis.

To examine the behavior of the conduction channels involved in these junctions, we followed the example of Vardimon *et al.* to calculate the transmittances of channels based on measured conductance and Fano factor [14]. Treating the junctions within the Landauer-Büttiker approach as coherent quantum conductors with transverse dimensions on the order of the electron Fermi wavelength, the conductance of an individual junction is described as the sum of independent conduction channels. The Fano factor is experimentally determined from the dependence of shot noise on applied bias and the conductance is given by the lock-in measurement. Therefore, given specific values for a paired $G$ and $F$, measured for a single junction or as an ensemble average of a dataset at a given applied bias, it is possible to calculate an array of all possible combinations of transmission coefficients that are compatible with the desired $G$ and $F$ values. The general assumption is that the group of coefficients must satisfy the following inequalities:

$$G - \Delta G \le G_0 \sum_{i=1}^{N} \tau_i \le G + \Delta G \text{ and } F - \Delta F \le \frac{\sum_{i=1}^{N} \tau_i(1-\tau_i)}{\sum_{i=1}^{N} \tau_i} \le F + \Delta F,$$

where $\Delta G = 0.01\ G_0$ (based on the bin size) and $\Delta F = \frac{\sigma}{\sqrt{N}}$, the standard error of the ensemble averaged Fano factor at each conductance. This process operates under the additional assumption $\tau_i \ge \tau_{i+1}$. A value of $\Delta\tau = 0.005$ is used as the bin size for transmission coefficients to ensure a finite number of solutions. This model is valid in the limit that the transmittance of the last channel goes to zero. The number of channels is limited to six for conductances up to 3 $G_0$, and at low bias, the last three channels all tend toward zero. The choice of six channels is reasonable for the low-conductance regime of interest because the number of channels for single atom contact is limited by the number of atomic valence orbitals [30]. Gold is a monovalent metal, so single atom conduction is carried out by one current-carrying channel, namely the *s* valence orbital, such that each gold atom can ordinarily contribute only up to one channel per 1 $G_0$ of conductance [14,30]. Thus in the limit of G < 3 $G_0$, we only expect the involvement of up to three conduction channels.

Histograms are created via selection of each set of six values for the transmission probabilities that combine to produce the desired pair of *G* and *F*. For an applied bias of 160 mV, Figure 2(c) displays an example histogram of calculated transmission coefficients for six conduction channels for *G* = 1.08, 2.18, and 3.07 $G_0$, with measured Fano factors of *F* = 0.07, 0.09, and 0.09, respectively. These results are quantitatively consistent with the work by Vardimon *et al.* [14] on Au junctions at 4.2 K. Additionally, like the previous findings of Bürki *et al.* and Vardimon *et al.*, we see single channel saturation up to 1 $G_0$, but at higher conductance values, several partially opened channels can be found [12-14]. This behavior follows the model of disorder in the presence of free electrons at a smooth constriction [12,13]. At higher conductances, we find that the Fano factor saturates above the theoretical single-partially-open-channel curve. Therefore we limit comparisons between low bias and high bias transmittance distributions to below 3 $G_0$.

*B. Bias dependence of noise*

As in the previous room temperature studies, we observed a superlinear dependence of measured noise on scaled bias at high applied bias voltages. In this high bias regime, the Fano factor was found to rise considerably above the theoretical minimum Fano factor line. Figure 3 shows examples of the progression of the ensemble averaged Fano factor rising above the theoretical minimum line as applied bias was increased. This enhancement was independent of the order in which the biases were applied, suggesting the effect is intrinsic and not a result of changes in the junction during data acquisition at different biases. Additionally, the junction was periodically annealed at currents of 2-5 mA in order to minimize any effects of work hardening or contaminants [31]. (In the main text, we have plotted the noise in terms of Fano factor to facilitate comparison with the no-channel-mixing scenario. In the Supporting Material, we show plots of the excess noise as a function of the bias voltage for representative conductances.)

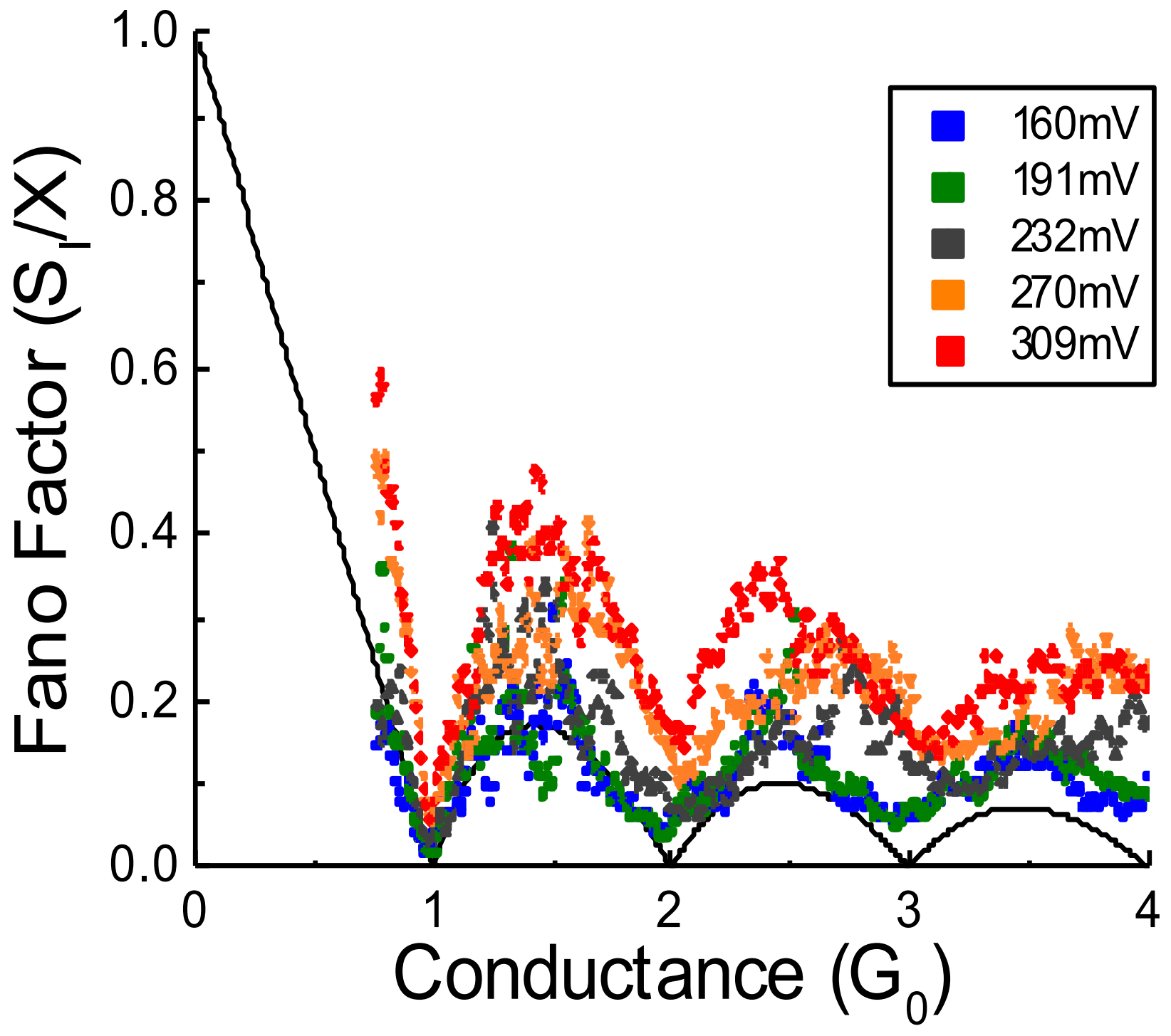

FIG. 3. Fano factor versus conductance as a function of applied bias across the junction and series resistance. Minimum Fano factor line is shown in black.

Figures 4(a) and 4(b) give the ensemble averaged Fano factor as a function of bias voltage across the contact for several conductance values, chosen to have similar Fano factors on the theoretical sequential channel opening line (shown in inset). While the trend shows some conductance-dependence, with spread increasing with increasing bias, some degree of enhancement appears to be universal across all conductances. In general, the enhancement factor appears to be even larger at 77 K than that seen at room temperature. For example, at 1.08 $G_0$ (Figure 4a) we observe a roughly four-fold enhancement of the Fano factor, compared to a less than twofold enhancement around the same conductance over similar bias range at room temperature.

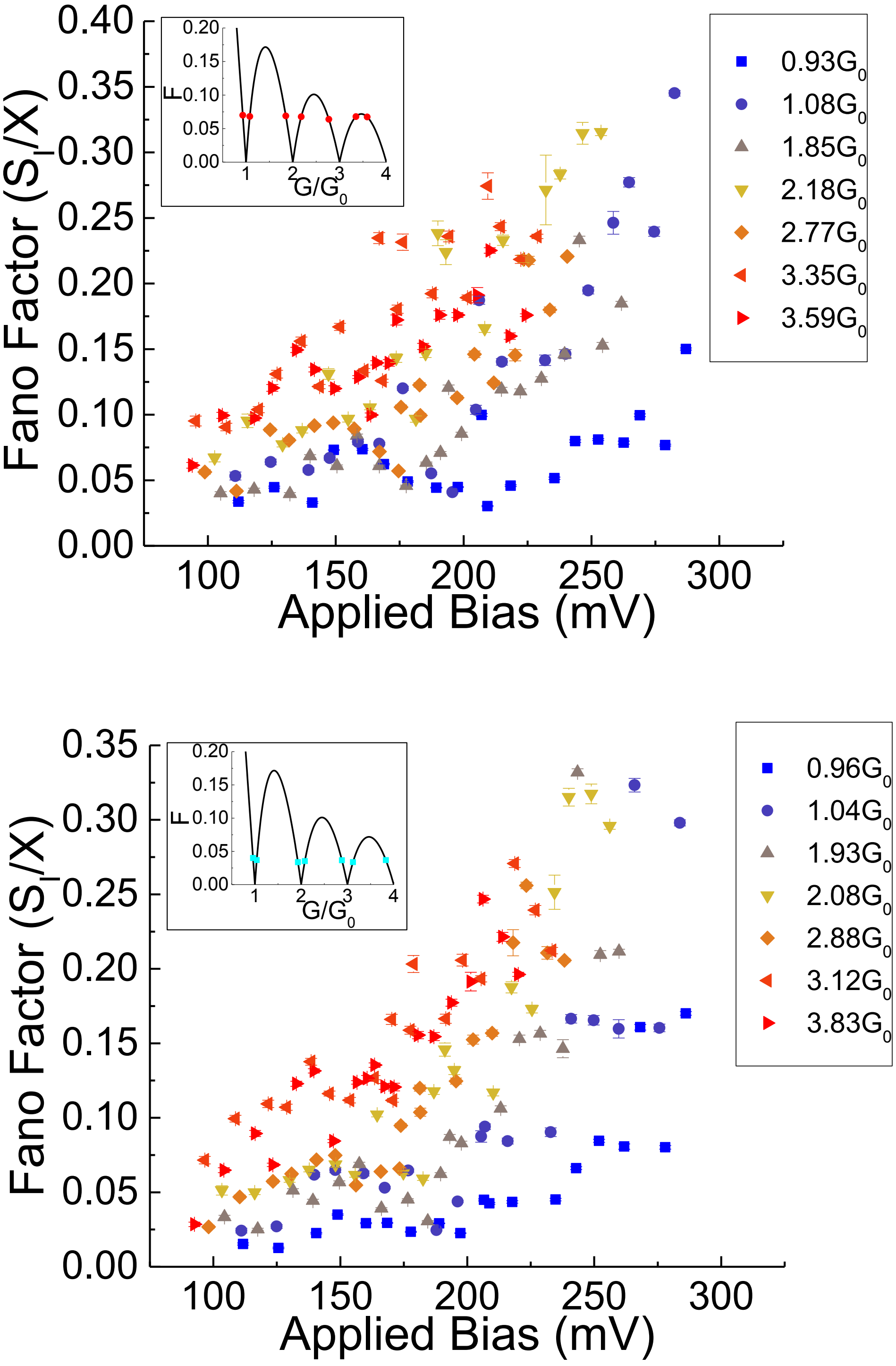

FIG. 4. Fano factor as a function of bias for conductances with similar values on the minimum Fano factor line (see points indicated on inset plots of no-channel-mixing Fano factor as a function of conductance). Note applied bias refers to the actual bias across the junction, taking into account the voltage dropped across the series resistance, $V_{appl} = V\frac{12906}{12906+(R_{stand}G)}$, where $G$ is expressed in units of $G_0$.

## IV. Discussion

A number of mechanisms could *a priori* lead to noise enhancements at large biases. As mentioned in prior work [15], extrinsic effects are quantitatively unlikely to be responsible for these observations. Flicker (*1/f*) noise has been ruled out previously by the observed bias dependence and by repeating measurements over a higher frequency bandwidth with no appreciable change in the observations. Similarly, trivial resistive heating of the bulk electrodes is ruled out. It is true that heating of the bulk electrodes when the junction is biased can increase the level of Johnson-Nyquist noise, and that perceived increase would appear in the measured lock-in signal. This effect, however, is set by the magnitude of the dissipated power and the thermal conductivity of the tip and metal films and was estimated at our experimental conditions to be far too small to explain the magnitude of the observed noise increase.

After ruling out these possibilities, we considered three candidate explanations for the nonlinear increase of noise at high bias: 1) the model of electronic heating as in Ref. 15; 2) increased noise via the bias-dependence of channel transmittances; and 3) modifications to the current noise due to interactions between electrons and local phonons [18,19,32-37]. Intrinsic electronic heating due to Fermi liquid viscosity effects would be expected to have comparatively weak temperature dependence, since the Au Fermi temperature (~64,000 K) is always much larger than experimentally accessible temperatures. Due to the observed conductance dependence, attempts to interpret the data as a rise in local effective electron temperature from the measured noise and Fano factor using the method of Chen *et al.* [15] were inconsistent with the previous results. The very simple model assumed local heating proportional to $IV$ and dominant thermal transport given by the electronic thermal conductivity calculated via the Wiedemann-Franz law, so that the locally limited increase of the electronic temperature (the width of the electronic distribution functions going into the junction) would be approximately independent of $G$.

Given that the increase in excess noise exhibited by our data does appear to depend on *G* and is also larger at low temperatures, we abandon local electronic heating as a candidate mechanism and look to other possible sources of noise enhancement.

We then considered the possibility that the observed Fano factor enhancement arises because the distribution of conductance channel transmittances, $\tau_i$, is changing with bias. In other words higher biases could alter the combination of $\tau_i$ that typically correspond to each conductance according to the minimal channel mixing picture. Changes in the preferred stable junction geometry arise from the large current densities and electric fields in the high bias regime [38]. At the largest biases applied in this work, the nominal current density is comparable to that employed in electromigration techniques [39]. It is conceivable then, that in response to large current densities, gold atoms in the junction may rearrange and acquire different stable configurations that do not necessarily favor fully transmitting channels.

If this effect were dominant, one would expect to find a strong bias dependence of conductance histograms, such as increased peak width or appearance of new or different conductance maxima for data acquired at high bias, reflecting a preference for junction geometries that favor increased mixing. Some signs of altered junction stability are observed in Figure 5. The overall number of counts for high bias data is lower despite the same number of breaking traces acquired because the traditionally stable junction configurations are less so at high bias. At low bias, a typical 1 $G_0$ conductance plateau during breaking lasted on the order of 100 ms, while at high bias, this lifetime dropped to 10 ms. At sufficiently large current densities, momentum may be transferred from the conduction electrons to the atoms at surfaces or grain boundaries, causing them to be displaced. In addition to the large current densities, high electric fields due to large potential differences over the short length scale of the junction may also contribute to mechanical instability. Therefore some configurations that are very stable at low biases are significantly less so at high bias, resulting in smaller peaks in the histograms. The histogram maxima are also asymmetric at high bias, developing shoulders toward lower conductances (Figure 5 inset), and we observe several additional peaks of smaller intensity, indicating a larger number of relatively stable configurations at non-integer conductance values. The new stable configurations, however, are relatively less frequent compared to the integer $G_0$ maxima and are not consistent between consecutive bias values. Overall, the general shape of the histograms stays qualitatively similar. It remains an open question then, whether dramatic changes in the transmittances can occur while maintaining the expected histogram forms. Alternatively, the large current densities could also contribute to inelastic processes in the junction, which may correlate to decreasing stability of junction configurations and thus the decreasing peak heights.

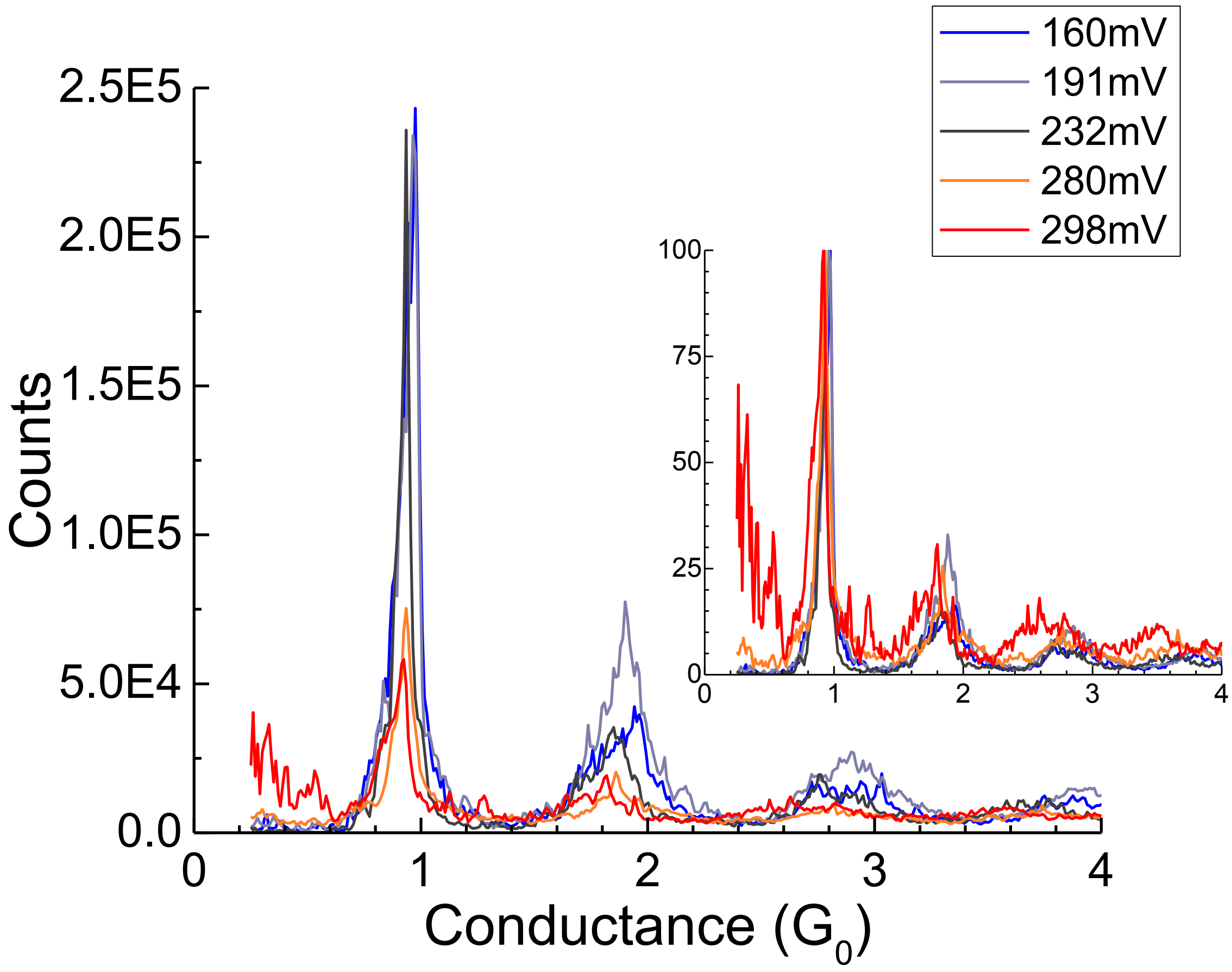

FIG. 5. Conductance histograms for increasing applied bias. The main plot demonstrates the decreasing number of stable junction configurations achieved at high biases. The inset histograms were normalized

to the 1 $G_0$ peak to emphasize the changes in shape as bias is increased. At high biases, we note a widening of the $nG_0$ peaks in addition to new peaks at non-integer conductances.

If bias-dependent transmittances are to explain the enhanced Fano factor, the $\tau_i$ would need to change drastically as channel mixing radically increases with bias. Figure 6 presents the inferred transmittance histograms for $G$ = 0.8, 1.08, and 2.18 $G_0$ for a range of applied bias voltages. We observe a universal *lowering* of the primary transmission channel and a broadening of all channels with increasing bias. According to the Landauer-Büttiker expression for shot noise at finite temperature, noise should be maximized when all channels contribute equally. Thus, the result that an enhanced Fano factor can be explained by greater contributions from more channels is not surprising. Therefore, we note that it is mathematically possible for channel mixing to account for the enhancement of the Fano factor. Bearing in mind the real-space origins of the conduction channels, as bias is increased, it is reasonable that tunneling conduction (and hence transmission) from one side of the junction to the other would be relatively enhanced compared to the low bias situation, favoring increased transmission for (small transmittance at low bias) tunneling channels. However, the physical process by which channels that are nearly fully transmitting at low bias would have *decreasing* transmission at high bias (necessary to satisfy the $G$ and noise constraints) remains unclear.

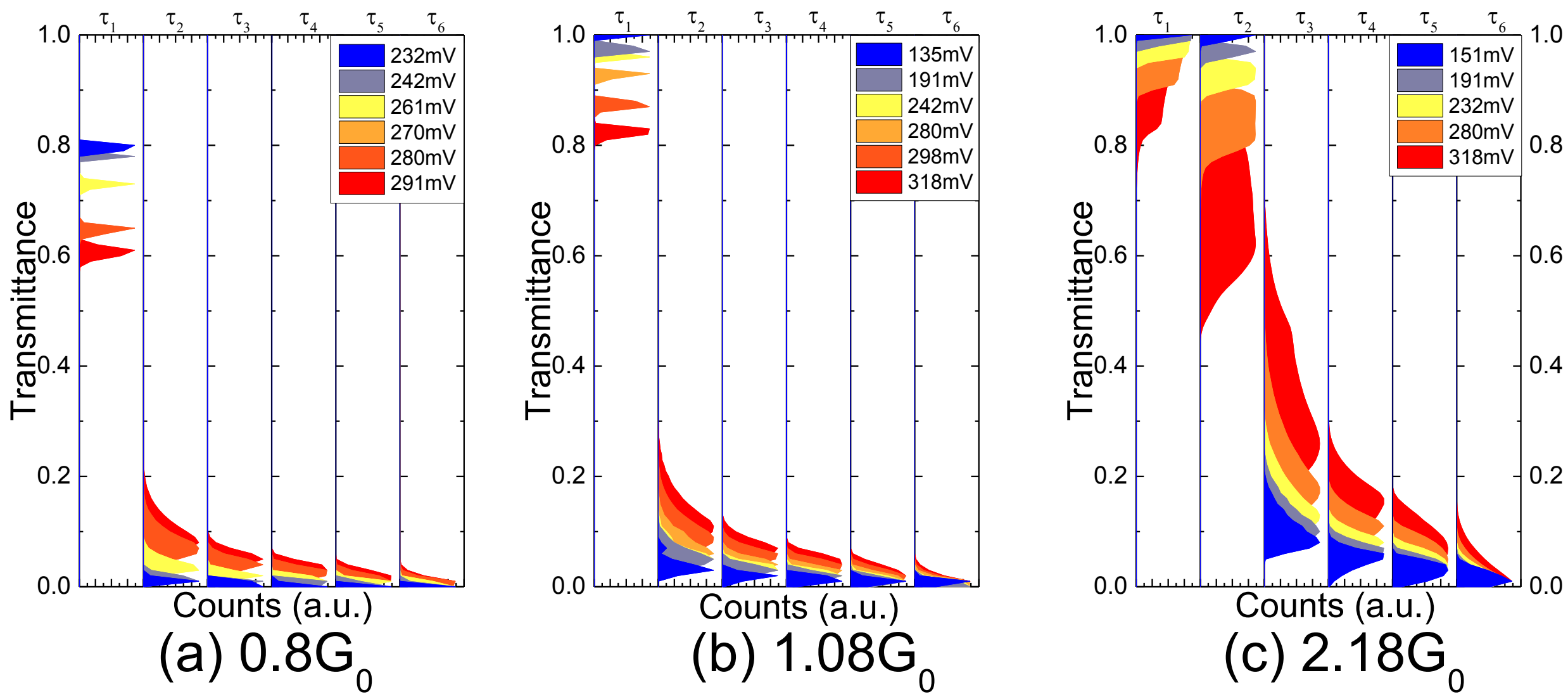

FIG. 6. Transmittance histograms inferred for increasing bias at **(a)** 0.8 $G_0$, **(b)** 1.08 $G_0$, and **(c)** 2.18 $G_0$ assuming that all of the increase in noise observed at high bias results purely from channel mixing. Note that various conductances would have to exhibit rather different evolutions of transmittance distributions in order to obtain fortuitously similar bias evolutions of Fano factor.

As another alternative, we also investigated the possibility of electrons coupling to a local vibrational mode. There have been many theoretical treatments of this general situation [18,19,32-37], with the precise predictions for the noise depending on several parameters such as the magnitude of the electron-vibrational coupling; multiple energy scales including the bias $eV$, the vibrational quantum $\hbar\omega$, and the broadening of the electronic states, $\Gamma$; and the coupling of the local vibrational mode to the bulk phonons. The theoretical predictions have largely been performed assuming single electronic channels and without considering ensemble averaging. Two limiting cases are particularly relevant: 1) thermal phonon populations, assuming relaxation between the local vibrational mode and the bulk phonons is fast, so that the local vibrational mode population is assumed to be thermally distributed at the temperature of the substrate, and 2) strongly nonequilibrium vibrational populations, such that the local vibrational mode's population is strongly athermal and predominantly driven by coupling to conduction electrons with energies sufficient to excite the mode.

In general, it is expected that the Fano factor will be altered at biases greater than the characteristic energy of the phonon mode [18,19,32-37]. This modification of $F$ depends on both the total transmittance of the junction [18,19,22,34,35] and on the coupling of the conductance channels to the phonon degree of freedom. In previous experiments looking at individual single-channel Au point contacts at cryogenic temperatures and low biases [22], discrete changes in $F$ at biases near the Au optical phonon energy are observed, with either increases or decreases in $F$ with increasing bias depending on the transmittance of the single dominant channel. In individual electromigrated Au junctions with multiple channels, qualitatively similar features are seen with higher threshold energies [39].

We chose to perform the present experiments at 77 K ($k_BT \sim 6.6$ meV), so that the condition $k_BT < \hbar\omega$ is fulfilled for the Au optical phonons, which have a characteristic energy of 17 meV. Therefore at this environmental temperature, thermal phonon populations should be greatly reduced relative to room temperature ($k_BT \sim 26$ meV). If the previously reported 300 K bias-dependence of $F$ resulted primarily from coupling to *thermalized* phonon populations, then one might expect considerably lower enhancement at 77 K. The observed persistence of the enhanced Fano factor in this low temperature regime appears to disfavor inelastic coupling to equilibrated optical phonons as a candidate mechanism for the growth of $F$ with bias at 77 K.

It has been theorized that a *nonequilibrium* phonon distribution enhances the Fano factor in the single-channel case that has been considered. Current density, passing through the atomic scale junction in the high bias limit is large and could excite athermal phonon populations. Theoretical models that assume strong pumping of the local phonon populations by the electrons predict contributions to the noise with voltage dependences of $V^3$ and $V^4$ [18,19,37]. The single transmitting channel model proposed by Novotný *et al*. [18,19] argues the $k$th moment of the inelastic current will be dependent on voltage as $V^{2k}$. Therefore, the current noise (the second moment of the current) should increase as $V^4$ when the primary inelastic process of the conduction elections is interactions with nonequilibrium phonons, though no precise calculation has been performed for the multichannel and ensemble-averaged situation.

Furthermore, this model contrasts the expected bias dependence for coupling to a phonon with a fixed average population (such as the thermally populated optical phonons) versus backaction of current-driven fluctuations of the phononic populations. While the dependence of current noise on bias voltage in the presence of a thermally equilibrated local phonon population is expected to be at most $V^2$, in the limit $eV > \hbar\omega$ and nonequilibrated phonon populations, the dependence of the noise is expected to be $V^4$. Measured excess current noise as a function of applied bias for five conductances is shown in the Supplementary Materials, as well as plots of the measured current noise as a function of $V^n$ for $n$=1-4. Because the variation within the ensemble average is sufficiently large it is difficult to make strongly constrained fits to precise bias dependences. It is possible, however, to plot the excess noise as a function of $V^n$ to identify the dominant power law in the high bias limit. Doing so, we seem to find increased linearity as $n$ approaches 4 in the high bias regime across the conductance range.

According to the model that takes into account strong pumping of the local phonon populations [18,19,32-37], at higher temperatures, one would expect the equilibrium contribution to dominate, while at cryogenic temperatures, the non-equilibrium, fluctuating populations should be the primary mechanism. Despite increases in shot noise above what is expected, the degrees of Fano factor enhancement were much smaller for the room temperature case than those observed in the present work. For example, at room temperature, the increase in Fano factor for 3 $G_0$ was from 0.35 at low bias to about 0.45 at about 225 mV, which constitutes a roughly 30% increase. In our observations at 77 K for a conductance of 3.12, the inferred Fano factor increased from about 0.08 at 100 mV to 0.28 around 225 mV, a 250% enhancement. The stronger bias dependence at 77 K versus room temperature may agree with coupling to nonequilibrated phonon populations as a good candidate mechanism for the enhanced Fano factor at high bias.

While this model describes the data relatively well, it must still be noted that it is constructed assuming a single conduction channel, which is distinct from and simpler than our situation of potentially multiple channels and ensemble averages over many junction configurations. While the multi-channel case should display the same asymptotic behavior as the single channel case [18], the theoretical treatment of ensemble averages is more complicated and has yet to be considered in full.

Another aspect to consider is the limit of the weak electron-phonon coupling theory, under which this model is developed. In other words, under what circumstances must the coupling be considered strong. In previous studies of atomic scale gold junctions, the *I-V* characteristics appeared Ohmic even at high bias, suggesting only small inelastic contributions to the current. Fitting our current noise data at 0.8 $G_0$ (best approximation to single-channel transport) to the model set forth by Novotný *et al.* [19], we calculate a dimensionless coupling constant of $\gamma_{\text{e-ph}} \approx 0.002$. Typically electron-phonon coupling is considered weak for $\gamma_{\text{e-ph}} < 0.1$, so the system appears to be within the limit. The bias dependence of noise in the strong coupling limit, however, has not yet been thoroughly calculated.

Experimentally, studies of flexible mechanical break junctions, in which junctions may be held at specific conductances, would be beneficial in obtaining measurements of both current and noise while bias is swept into the high bias limit, permitting a direct comparison between the bias dependence of first two moments of current. In these systems, clear evidence of the evolution of current noise with bias in individual junctions could illuminate whether the enhancement of noise follows the predicted bias dependence for phonon contributions. The conductance dependence may also be clarified by comparing junctions of different configurations to determine if noise increases for all conductances or rather, like seen in Ref. 22, decreases at certain transmittances. A study of this nature is necessary to firmly establish the nature of the observed enhancement of noise in the high bias limit.

## V. Conclusions

We have presented the bias dependence of the ensemble averaged Fano factor in STM-style Au junctions. While at low biases, the Fano factor behaves as expected based on Landauer- Büttiker formalism, as bias is increased, the Fano factor is enhanced above the theoretical minimum line. Ruling out mechanisms such as flicker noise or bulk heating, and finding that the previously attempted model of electronic heating is inconsistent with the observed conductance dependence, we consider two candidate processes as the source of this increased noise: 1) increased channel mixing at high bias; and 2) back-action by current-driven phonon populations. Current densities on the order of those employed in electromigration techniques could also be responsible for rearrangement of the junction atoms such that different geometries are stable at higher biases. A reordering of the junction atoms would result in more channels being involved in conduction than is typically predicted for each conductance. While we are able to computationally show how this could result in an enhancement of the Fano factor, we also note that the conductance histograms still display preference for the fully transmitting channels. Large current densities at high biases in atomic-scale junctions could excite nonequilibrium phonon populations, which in turn inelastically scatter conduction electrons. In the single-channel, single junction configuration case, inelastic backaction of these fluctuating phonon populations is expected to increase the noise and to have a stronger dependence on bias than interactions with equilibrated phonons. While that agrees with the higher enhancement observed here at temperatures below the gold optical phonon energy, we point out that a full theoretical analysis with multiple channels and considering the ensemble average is needed. Experimentally, future work in mechanical break junctions could investigate the inelastic limits of atomic junctions by simultaneously measuring the first two moments of the current, and further theoretical modeling may clarify the relationship between the most likely transmission coefficients and the conductance histograms for large ensembles of data.

The authors acknowledge supported by NSF DMR-1305879.

**Supplementary Materials**

**Current noise enhancement: channel mixing and possible nonequilibrium phonon backaction in atomic-scale Au junctions**

Loah A. Stevens[1], Pavlo Zolotavin[1], Ruoyu Chen[1], and Douglas Natelson[1,2,3]

[1]*Department of Physics and Astronomy, Rice University, 6100 Main St., Houston, TX 77005*

[2]*Department of Electrical and Computer Engineering, Rice University, 6100 Main St., Houston, TX 77005*

[3]*Department of Materials Science and NanoEngineering, Rice University, 6100 Main St., Houston, TX 77005*

Based on various theoretical analyses of the excess current noise in atomic-scale junctions in the presence of local phonon modes, we are motivated to consider a dependence of the excess current noise on bias of the functional form $S_I(V)-S_I(0)=a+bV+cV^2+dV^4$. Linear dependence of current noise on bias is the expected result for the Landauer-Büttiker picture. Quadratic and $V^4$ terms can result from interactions between electrons and equilibrated or unequilibrated phonon populations, respectively [S1,S2]. The scatter in the ensemble-averaged data make it very challenging to constrain polynomial fit coefficients. As an alternative, to try to identify the dominant power law at high bias, we plotted the measured excess noise as a function of $V^n$ for $n$=1-4 for several conductances. An apparent linear dependence vs. $V^n$ would suggest dominance of that power law, at least in the high bias limit. The results for 0.8, 1.08, 1.85, 2.18, and 2.88 $G_0$ are shown in the figures below. Analysis across these representative conductance values indicates reasonable consistency with a high bias limiting dependence of $n$=4. This is true for conductances across the range from 0.8 to 3 $G_0$.

We would like to point out that the correspondence of the high bias voltage dependence of $S_I(V) - S_I(0)$ with the functional form predicted in Ref. S1,S2 for unequilibrated phonon populations should be considered only as indirect evidence in favor of this mechanism for noise enhancement. Analysis of the individual noise *v.* bias traces similar to Ref. 22 of the main text would be necessary for a stronger statement as opposed to the ensemble averaged results presented here. Alternatively, a theoretical modeling that incorporates ensemble averaging with high bias limit voltage dependence of current noise could help clarify this issue.

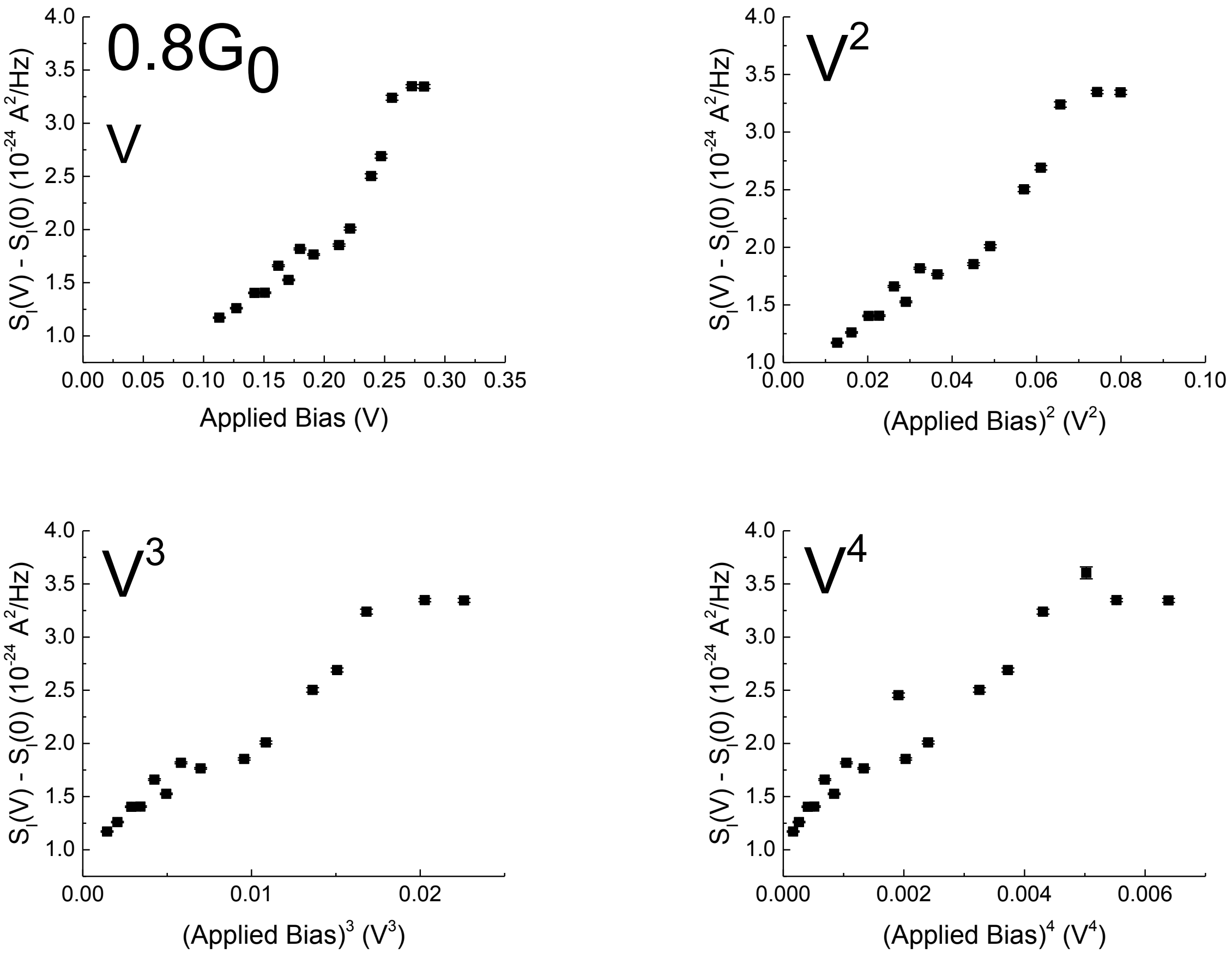

**Fig. S1.** Ensemble-averaged excess noise as a function of powers of voltage bias for junctions with a conductance of 0.8 $G_0$.

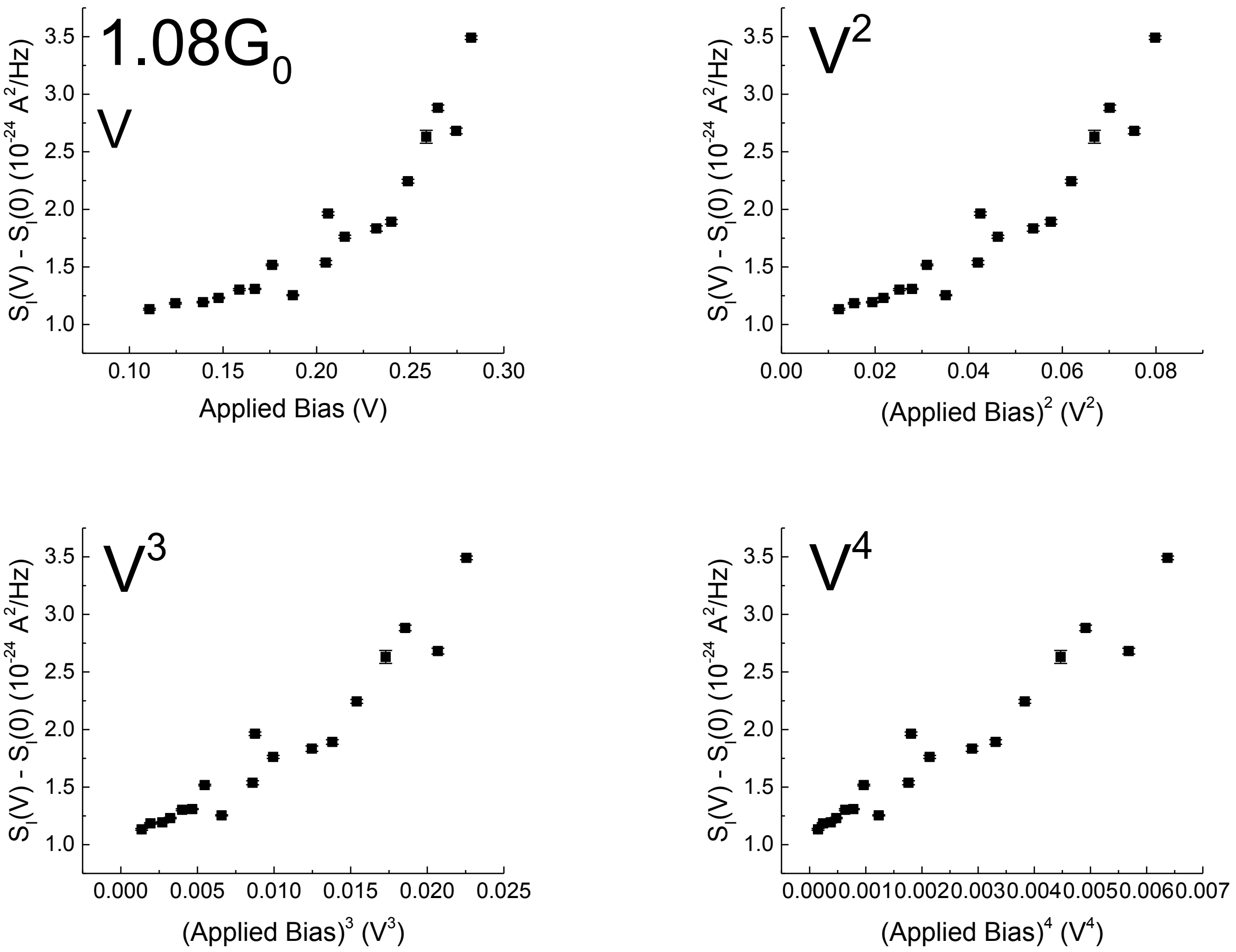

**Fig. S2.** Ensemble-averaged excess noise as a function of powers of voltage bias for junctions with a conductance of 1.08 $G_0$.

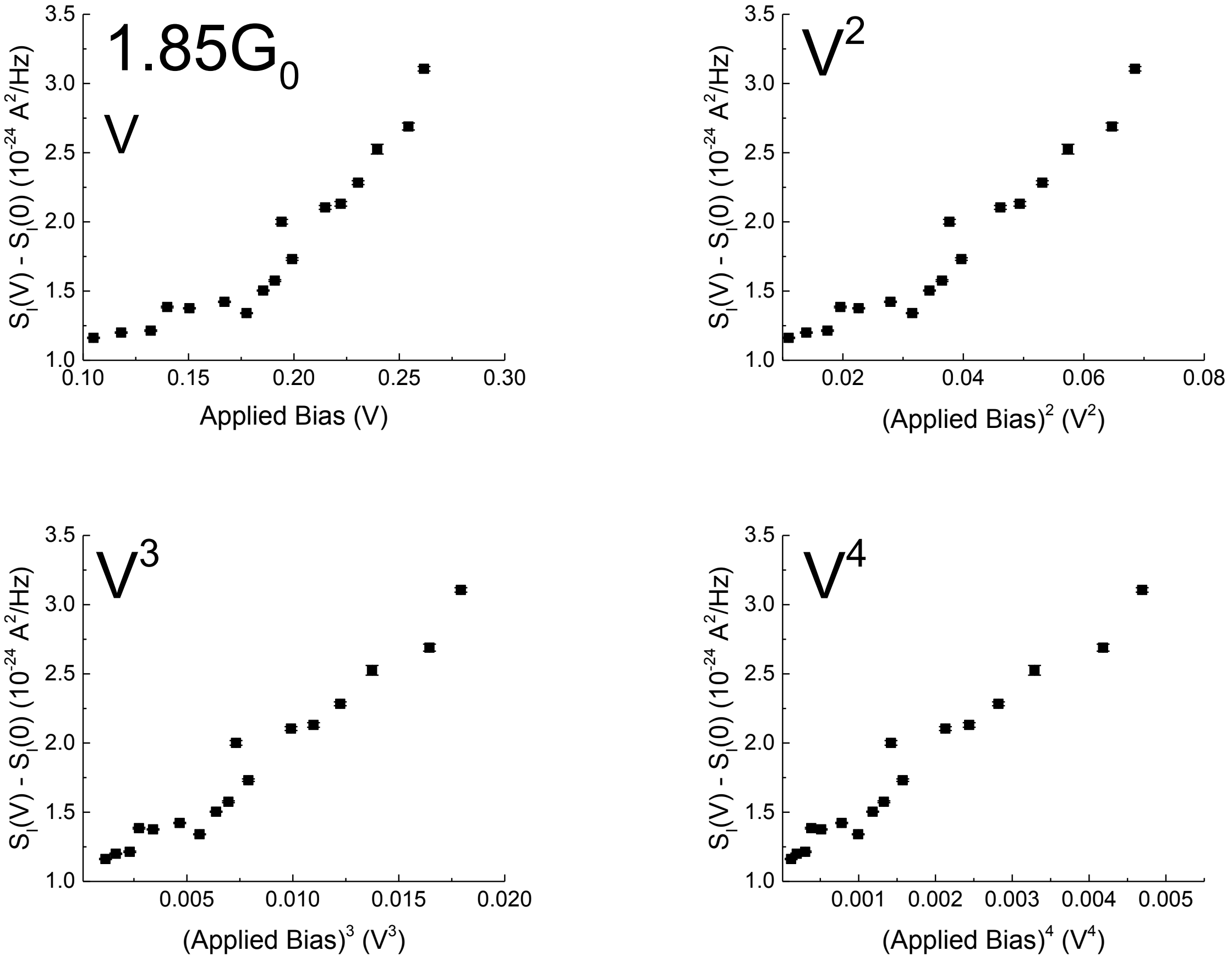

**Fig. S3.** Ensemble-averaged excess noise as a function of powers of voltage bias for junctions with a conductance of 1.85 $G_0$.

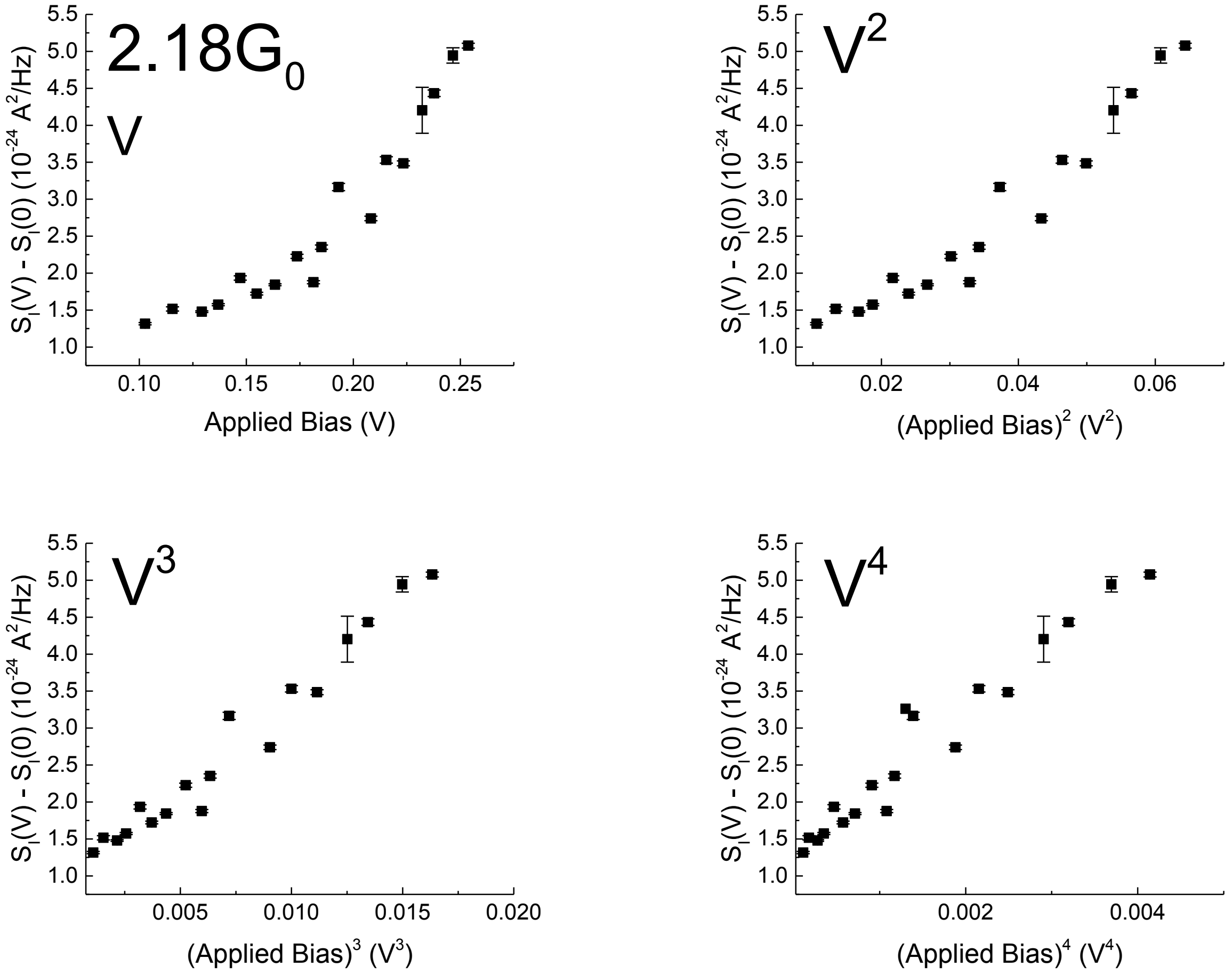

**Fig. S4.** Ensemble-averaged excess noise as a function of powers of voltage bias for junctions with a conductance of 2.18 $G_0$.

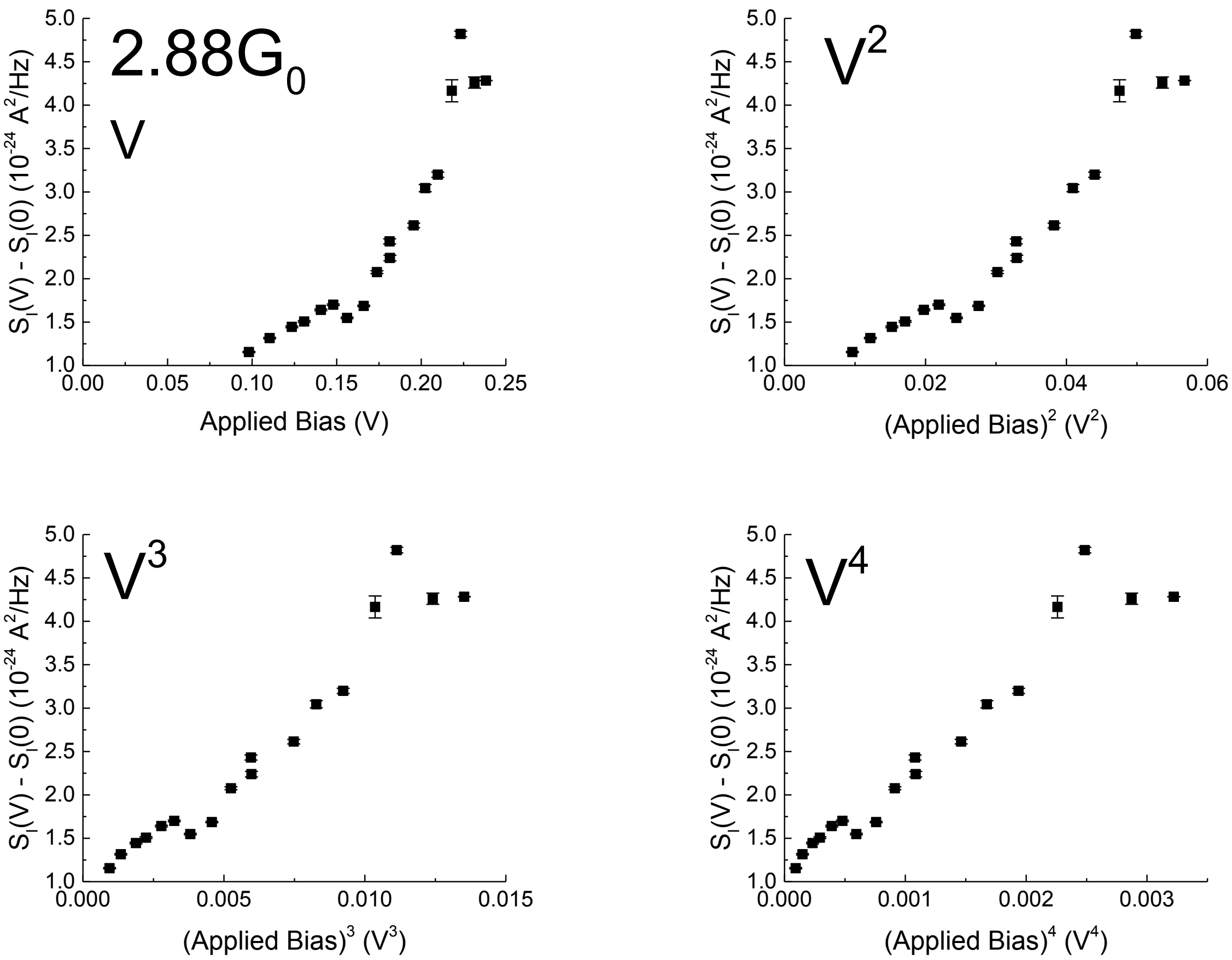

**Fig. S5.** Ensemble-averaged excess noise as a function of powers of voltage bias for junctions with a conductance of 2.88 $G_0$.